\documentclass[preprint]{elsarticle}
\usepackage[utf8]{inputenc}

\usepackage{etoolbox}

\usepackage{amsmath}
\usepackage{amssymb}
\usepackage{float}
\usepackage[bookmarks=false]{hyperref}
\nopreprintlinetrue

\begin{document}
\begin{frontmatter}

\title{The Inferred Abundance of Interstellar Objects of Technological Origin}
\author[1]{Carson Ezell\corref{cor1}}
\ead{carson.ezell@cfa.harvard.edu}
\author[1]{Abraham Loeb}
\ead{aloeb@cfa.harvard.edu}

\address[1]{Department of Astronomy, Harvard University, 60 Garden St, 02138, Cambridge, MA, USA}

\cortext[cor1]{Corresponding author}

\date{September 2022}

\begin{abstract}
The local detection rate of interstellar objects can allow for estimations of the total number of similar objects bound by the Milky Way thin disk. If interstellar objects of artificial origin are discovered, the estimated total number of objects can be lower by a factor of about $10^{16}$ if they target the habitable zone around the Sun. We propose a model for calculating the quantity of natural or artificial interstellar objects of interest based on the object’s velocity and observed density. We then apply the model to the case of chemically propelled rockets from extraterrestrial civilizations. Finally, we apply the model to three previously discovered interstellar objects— the object ‘Oumuamua of unknown origin and the first interstellar meteors CNEOS 2014-01-08 and CNEOS 2017-03-09. 

\end{abstract}

\end{frontmatter}

\section{Introduction}

Recent surveys have allowed for the detection of the first four known interstellar objects over the past decade: the interstellar meteors CNEOS 2014-01-08 \cite{siraj_2019_2022} and CNEOS 2017-03-09 \cite{siraj_interstellar_2022}, the interstellar object ‘Oumuamua \cite{meech_brief_2017, micheli_non-gravitational_2018} and the interstellar comet Borisov \cite{guzik_initial_2020}. The rate of detection of interstellar objects depends on detection sensitivity, given the objects’ size and distance. One can use recent rates of detection of interstellar objects and known capabilities to estimate the density of similar objects in the solar neighborhood and the total number of such objects bound by the thin disk of the Milky Way \cite{siraj_2019_2022}. 

Estimates for density and quantity of naturally occurring interstellar objects bound by the Milky Way’s thin disk assume that interstellar objects are initially ejected from their host stars in random directions \cite{siraj_identifying_2019}, and that they are vertically distributed with respect to the galactic plane based on the scale height of their parent stars. Such assumptions still hold if the objects are artificial and undirected, such as space debris from extraterrestrial technological civilizations (ETCs). However, assuming random directions of interstellar objects does not apply if the objects are artificial and directed towards regions of interest, such as the habitable zones of stars. Estimating the frequency and density of interstellar objects given they are targeted requires considering the quantity of targets and volume of targeted regions. 

By applying the Copernican Principle, which suggests that human civilization on Earth is a typical example of the emergence of intelligent life in the universe, we gain insights into possible targets of extraterrestrial technological artifacts (ETAs). The traditional Search for Extraterrestrial Intelligence (SETI), focused on electromagnetic signals \cite{drake_radio_1965} which escape the Milky Way galaxy and cannot be detected a light-crossing time after transmission ceases. Indeed, one of the critical parameters in the Drake equation is the lifetime of a transmitting civilization \cite{westby_astrobiological_2020}. However, spacecraft propelled by chemical propellants do not exceed the local escape speed for the Milky-Way galaxy, remain gravitationally bound to it and accumulate over time summing over the rise and fall of space-venturing civilizations.

In this paper, we develop a model for calculating the total number of interstellar objects of technological origin bound by the galactic thin disk based on observational data, and we find that the quantity of objects is significantly reduced if the objects’ trajectories are targeted. The outline of the paper is as follows. In Section 2 we calculate the expected velocities of interstellar objects given their launch characteristics. In Section 3, we calculate the local density and total quantity of interstellar objects bound to the Milky-Way given our detection rates. In Section 4, we consider the implications for the previously detected interstellar objects.

\section{Launch Characteristics}

First, let us consider the motion of an interstellar object of interest ejected from a star system. An object ejected from a planet, or other celestial body in orbit around a central star, must escape from the gravitational well of both the planet and the central star, ultimately achieving a velocity $v_{\infty}$ at a large distance. The escape velocity from a planet is given as, \begin{equation} v_{esc}^p= \sqrt{\frac{2GM}{R}} \text{ ,} \end{equation}
where $G$ is the gravitational constant, $M$ is the mass of the planet, and $R$ is the radius of the planet. The escape velocity from a star system is given as, \begin{equation} v_{esc}^s=\sqrt{\frac{2GM_{\star}}{a_{\star}}}\text{ ,} \end{equation}
where $M_{\star}$ is the mass of the central star and $a_{\star}$ is the orbital radius of the planet around the star. 

Assuming the object’s initial velocity is in the same direction as the circular motion of the planet around the central star, it shares the same velocity of the planet, 
\begin{equation} v_{circ} = \sqrt{\frac{GM_{\star}}{a}}\text{ .} \end{equation}
Hence, for an object ejected from the surface of a planet, its necessary velocity increment to escape from both the gravitational well of the planet and the gravitational well of the central star is,
\begin{equation} \Delta v = \sqrt{(v_{esc}^p)^2 + (\frac{1}{2})(v_{esc}^s)^2} \text{ .} \end{equation} 

Assume an object is ejected with an initial velocity relative to the surface of the planet of $v_i$ where $v_i > v_{esc}^p$. Then, after escaping the gravitational well of the planet, we have a velocity relative to the central star of, 
\begin{equation} v_s \approx v_{circ}+\sqrt{v_i^2-(v_{esc}^p)^2}\text{ .} \end{equation}
If $v_s > v_{esc}^s$, it follows that, 
\begin{equation} v_{\infty} \approx \sqrt{(v_s)^2-(v_{esc}^s)^2}\text{ .} \end{equation}
Given the exit velocity of an object of interest from its origin star, we consider the mean velocity dispersion of all such objects bound by the thin disk of the Milky Way galaxy \cite{binney_disk_2008}. 

The mean velocities of stars in the thin disk of the Milky Way are given as $(V_R, V_{\Phi}, V_Z) = (-1, -239, 0) \pm (0.28, 0.21, 0.44) \text{ km s}^{\text{-1}}$, and the velocity dispersions of stars in the thin disk of the Milky Way are $(\sigma_R, \sigma_{\Phi}, \sigma_Z)=(31, 20, 11) \pm (0.24, 0.17, 0.6) \text{ km s}^{\text{-1}}$ \cite{vieira_milky_2022}.

Let $\sigma_Z^i$ be the mean velocity dispersion of objects of interest in the vertical direction after escaping their parent star system. If objects are ejected in random directions, then,
\begin{equation} \sigma_Z^i=\frac{v_{\infty}}{\sqrt{3}}\text{ ,} \end{equation}
assuming an ensemble of objects all ejected at the same speed, or restricted to some velocity bin out of the entire population. 

For a given ensemble of objects labeled by the index $i$, with velocity relative to its star $v_{\infty}$, the vertical velocity dispersion is given as,
\begin{equation} \sqrt{(\sigma_Z)^2+(\sigma_Z^i)^2} \text{ .}\end{equation}
For civilizations that emerge in conditions similar to our own, we assume artificial objects to be launched from the habitable zones of star systems, where the proper conditions exist for a rocky planet to to present and maintain the proper temperature to support liquid water on its surface \cite{lingam_life_2021}. The stellar luminosity is given by, 
\begin{equation} L_{\star}=4 \pi R_{\star}^2\sigma_{SB}T_{\star}^4 \text{ ,} \end{equation}
where $T_{\star}$ is the temperature of the star, $R_{\star}$ is the radius of the star, and $\sigma_{SB}$ is the Stefan-Boltzmann constant. Assuming we are interested in planets that have a similar surface temperature to that of Earth, 
\begin{equation} \sigma_{SB}T_{eq}^4=\frac{L_{\star}}{16 \pi a_{\star}^2}(1-A) \text{ ,} \end{equation}
where $A$ is the albedo of the planet and $T_{eq}$ is the equilibrium temperature of the planet. If the planet has the same albedo as Earth, we have, 
\begin{equation} a_{\star} = 1 \text{ AU} (\frac{L_{\star}}{L_{\odot}})^{\frac{1}{2}} \text{ ,}\end{equation}
where $L_{\odot}$ is the luminosity of the Sun \cite{lingam_life_2021}. Using the luminosity-mass relation for common low-mass stars $L_{\star} \propto M_{\star}^3$ \cite{bohm-vitense_introduction_1992}, we have, 
\begin{equation} a_{\star} = 1 \text{ AU} (\frac{M_{\star}}{M_{\odot}})^{\frac{3}{2}}\text{ ,} \end{equation}
where $M_{\odot}$ is the mass of the Sun.
Using equation (2), as, 
\begin{equation} v_{esc}^s=v_{esc}^{\odot}(\frac{M_{\star}}{M_{\odot}})^{-\frac{1}{4}}\text{ ,}\end{equation}
where $v_{esc}^{\odot} = 42.1 \text{ km s}^{\text{-1}}$ is the escape velocity at the Earth-Sun separation. Using equation (1) and the mass-radius relationship $M \propto R^{3.7}$ for rocky planets \cite{zeng_mass-radius_2016}, \begin{equation} v_{esc}^p = v_{esc}^{\oplus}(\frac{R}{R_{\oplus}})^{1.35} \text{ ,}\end{equation}
where $R_{\oplus}$ is the radius of the Earth and $v_{esc}^{\oplus}=11.2 \text{ km s}^{\text{-1}}$ is the escape velocity from Earth’s surface. Thus, equation (4) yields, 
\begin{equation} \Delta v = \sqrt{(v_{esc}^{\oplus})^2(\frac{R}{R_{\oplus}})^{2.7} + \frac{1}{2}(v_{esc}^{\odot})^2(\frac{M_{\odot}}{M_{\star}})^\frac{1}{2}}\text{ .} \end{equation}
Invoking the Copernican Principle and considering the solar system to be typical of star systems where intelligent life emerges, we adopt mean values of $R=R_{\oplus}$ and $M_{\star}=M_{\odot}$ with $v_{esc}^{\oplus}=11.2 \text{ km s}^{\text{-1}}$ and $v_{esc}=42.1 \text{ km s}^{\text{-1}}$. We consider the example of chemically propelled probes as artificial interstellar objects from intelligent civilizations that we may observe, where $v_i \sim 30 \text{ km s}^{\text{-1}}$.  Equation (6) then yields  $v_{\infty}=39.31 \text{ km s}^{\text{-1}}$. Furthermore, using equations (7) and (8), the find the mean vertical velocity dispersion relative to the Local Standard of Rest (LSR) is $25.22 \text{ km s}^{\text{-1}}$ for chemically-propelled rockets.

\section{Local Density and Detection Rate}

Given the mean velocity $v_{\infty}$ of interstellar objects of interest, we can use the detection rate in a given region to estimate the object density. Let $r_d$ be the detection rate of the objects, and let $A_d$ be the effective cross-sectional area of the survey region where detection is possible. For any class of objects with velocity $v_{\infty}$ relative to the Sun, we can calculate the number density of such objects in the solar neighborhood as,  
\begin{equation} \rho_i^{\odot} \approx \frac{r_d}{v_{\infty}A_d}\text{ .} \end{equation}
The estimated number density of objects bound by the thin disk is further affected by the scale height of the objects. The density estimate for objects at height $z$ above the Galactic plane is,
\begin{equation} \rho_i^z = \rho_0e^{-\frac{\phi_z}{\sigma_z^2}}\text{ ,} \end{equation}
where $\rho_0$ is the density of objects of interest in the galactic plane and $\phi_z$ is the gravitational potential at a vertical distance $z$ from the galactic plane.  We consider the case of a razor thin Galactic disk with local surface density, $\Sigma$, and a stellar scale height,
\begin{equation} h_{\star} \approx \frac{\sigma_z^2}{G\Sigma}\text{ .} \end{equation}
The relation between the scale height and vertical velocity dispersion is $\sigma^2 \propto h_z$. Thus, the mean scale height of objects under consideration is,
\begin{equation} h_i = h_{\star}(\frac{\sigma_z^i}{\sigma_z})^2\text{ .} \end{equation}
Then, the density of objects of interest at a given height $z$ is,
\begin{equation} \rho_i^z=\rho_0e^{-\frac{z}{h_i}} \text{ .}\end{equation}
\begin{figure}[H]%
\centering
\includegraphics[width=0.75\textwidth]{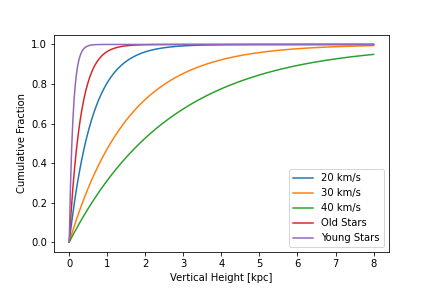}
\caption{Cumulative fraction of young stars, old stars \cite{binney_disk_2008}, and probes launched at different initial velocities bound by the Milky Way thin disk as a function of vertical height $z$ above the galactic plane}\label{fig1}
\end{figure}

We assume a local stellar number density of $\rho_{\star}$. If the objects are not targeted, we estimate the abundance of objects of interest per star in the solar neighborhood as $\frac{\rho_i^{\odot}}{\rho_{\star}}$. However, if interstellar objects are directed, we must consider their higher concentrations in targeted regions. We assume regions of interest are star systems or regions within star systems (e.g. habitable zones). 

For functional probes on targeted missions, we consider $V_t$ be the mean volume of a targeted region around a star. Then, the estimated local density of targeted objects in the solar neighborhood is reduced by a factor of $\sim \rho_{\star}V_t$. Targeted objects may be launched towards particular star systems, such as those which contain habitable planets. The estimated mean local density of objects is then smaller by another factor of $\sim p_t$, or the probability that a given planetary system is targeted because it hosts habitable rocky planets.

We adopt  $p_i=1$ if objects are directed, and $p_i=0$ otherwise. Based on our observations of object density in the solar neighborhood, we estimate the total quantity of objects of interest bound by the thin disk as,
\begin{equation} n_i = \int_0^{\infty} (\int_0^{\infty}(p_t\rho_tV_t)^{p_i}(\frac{r_d}{v_{\infty}A_d})e^{-\frac{(z-z_{\odot})}{h_i}}dz)2\pi r e^{-\frac{(r-r_{\odot})}{r_0}}dr \text{ ,}\end{equation}
where $z_{\odot}$ is the vertical distance of the Sun above the galactic plane, $r_{\odot}$ is the distance of the Sun from the galactic center along the midplane, and $r_0$ is the scale radius of the Milky Way. We adopt values of $z_{\odot}=20.8 \text{ pc}$ \cite{bennett_vertical_2019}, $r_0 = 3 \text{ kpc}$, and $r_{\odot} = 8 \text{ kpc}$ based on existing data \cite{binney_disk_2008}.

\begin{figure}[H]%
\centering
\includegraphics[width=0.75\textwidth]{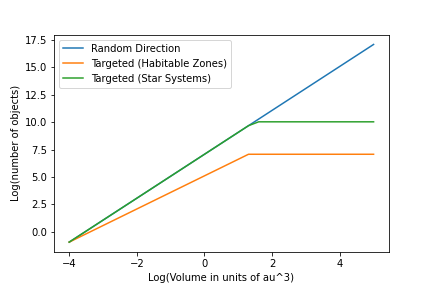}
\caption{Inferred abundance of technosignatures within a spherical survey region around the Sun as a function of its volume for random and targeted probes}\label{fig2}
\end{figure}

Suppose astronomical observations led to discoveries of chemically-propelled probes in the habitable zone of our solar system with an observed local density $\rho_c^{\odot}$. The local stellar number density is  $\rho_{\star} \approx 0.1 \text{ pc}^{\text{-3}}$ \cite{holmberg_local_2000}. For objects targeted towards habitable zones, we have $V_t \approx 1 \text{ AU}^{\text{3}}$ or $\approx 10^{-16} \text{ pc}^{\text{3}}$. Furthermore, a common estimate of the fraction of star systems with Earth-like planets in habitable zones is $p_t \sim 0.1$ \cite{lingam_life_2021}, and it is plausible that only these star systems would be targeted by probes searching for extraterrestrial life. 

Given the aforementioned properties of chemically-propelled rockets and a hypothetical detection rate of $\rho_c^{\odot} = 0.1 \text{ yr}^{\text{-1}}$ for interstellar meteors of meter size that collide with Earth, equation (21) estimates a total of $3.65 \times 10^{34}$ such objects bound by the Milky-Way thin disk if they are not targeted, or $3.65 \times 10^{18}$ objects if they are targeted.

\section{Discussion}

We estimate the total quantity of interstellar objects similar to those we have already detected, including the interstellar object of unknown origin ‘Oumuamua and the interstellar meteors CNEOS 2014-01-08 (IM1) and CNEOS 2017-03-09 (IM2).

‘Oumuamua has an estimated diameter of about 200 meters \cite{meech_brief_2017}. The Pan-STAARS telescope, which detected ‘Oumuamua, operated by conducting 6,600 “quad” (4 revisit) operations over 80 day periods, each period resulting in a detection volume of $0.3 \text{ AU}^{\text{-3}}$ for an object with the size and other properties of ‘Oumuamua  \cite{do_interstellar_2018}. This results in a detection volume of $\sim 1.37 \text{ AU}^{\text{3}} \text{ yr}^{\text{-1}}$. Given the capabilities of the Pan-STAARS survey and the lack of detections of further ‘Oumuamua-like objects, the abundance of ‘Oumuamua-like objects in the solar neighborhood is approximately $\rho_i^{\odot}=0.1 \text { AU}^{\text{-3}}$, or $10^{15} \text{ pc}^{\text{-3}}$
\cite{siraj_mass_2022}. Based on equation (21), our estimate for the total quantity of ‘Oumuamua-like objects bound by the thin disk  if they are not targeted is $\approx 1.91 \times 10^{26}$ objects, which aligns with previous estimates for the abundance of similar objects \cite{jewitt_interstellar_2022}. This estimate applies both in the case of ‘Oumuamua being of natural origin, and ‘Oumuamua being artificial space debris that is not targeted towards a particular location in space. 

However, the inferred abundance of probes is distinctly different in case of ‘Oumuamua-like objects being targeted towards particular regions of the galaxy, specifically habitable zones containing planets. ‘Oumuamua was detected at a distance of $\sim 0.2$ AU from Earth, and it passed through the habitable zone of our solar system \cite{desch_1ioumuamua_2021}. The estimated total number of ‘Oumuamua-like objects would then fall to $\approx 1.91 \times 10^{10}$.

The interstellar meteor IM1 had an estimated diameter of $\sim 0.45$ m and velocity of $60$ km $\text{s}^\text{-1}$, but it was detectable when it entered and burned up within the atmosphere of the Earth \cite{siraj_2019_2022}. The estimated detection rate for interstellar meteors similar to CNEOS is at least $\sim 0.1 \text{ yr}^{\text{-1}}$ \cite{siraj_2019_2022}, resulting in a local density estimate of $\rho_i^{\odot} \sim 10^6 \text{ AU}^{\text{-3}} = 10^{22} \text{ pc}^{\text{-3}}$. We then estimate $7.59 \times 10^{34}$ IM1-like objects bound by the thin disk of the Milky Way. However, if objects with the properties of IM1 were targeted towards habitable zones containing planets, we estimate $7.59 \times 10^{18}$ such objects. IM2 had a similar inferred number density to IM1 and a velocity of $40$ km $\text{s}^\text{-1}$ relative to the Local Standard of Rest \cite{siraj_interstellar_2022}. We estimate $2.78 \times 10^{34}$ IM2-like objects, and our estimate would be decreased to $2.78 \times 10^{18}$ if such objects were targeted towards habitable zones.

\section{Conclusion}

The abundance of ISOs depends on their size and can be calibrated through future surveys such as the Legacy Survey of Space and Time (LSST) on the Vera C. Rubin Observatory in Chile.\footnote{\href{https://www.lsst.org/about}{https://www.lsst.org/about}} Data from the James Webb Space Telescope may identify the nature and 3D trajectory of more 'Oumuamua-like or other interstellar objects crossing through or trapped within the solar system \cite{siraj_identifying_2019, hoover_population_2022}. 

The Galileo Project\footnote{\href{https://projects.iq.harvard.edu/galileo/home}{https://projects.iq.harvard.edu/galileo/home}} was established in 2021 \cite{loeb_overview_2022} as a scientific search for potential astro-archaeological artifacts from ETCs, including anomalous interstellar objects that may be revealed by the Webb telescope of Vera C. Rubin Observatory. The Galileo Project will also search for Unidentified Aerial Phenomena (UAP) within the atmosphere of the Earth \cite{loeb_overview_2022}. As objects of interstellar origin are discovered by the Galileo Project, number density estimates can be improved based on detection rates. If extraterrestrial equipment is discovered in a survey of space, we can estimate the total quantity of such objects conditional on whether they are defunct space debris or functional probes. 

Furthermore, the quantity of ISOs allows for calculations of the necessary mass budget per star to produce the objects. If interstellar objects originate in protoplanetary disks, our estimates for quantity can place further constraints on the fraction of mass that disks must eject to achieve the given population of objects \cite{siraj_preliminary_2021}. Analysis of ISOs can also provide insights into rare or unexpected phenomena, including an unusual material composition or whether they may transport prebiotic or biotic materials across interstellar distances \cite{siraj_new_2021}. The expedition planned by the Galileo Project to retrieve the interstellar fragments from CNEOS 2014-01-08 will be able to provide more insights into its rare material strength, isotope abundances, and nature \cite{siraj_ocean_2022}.

\section{Acknowledgements}

This work was supported in part by the Galileo Project at Harvard University.

\raggedright
\bibliography{main}

\end{document}